\documentclass[aps,pra,superscriptaddress]{revtex4}

\usepackage{graphicx}
\usepackage{amssymb}
\usepackage{amsmath}
\usepackage[retainorgcmds]{IEEEtrantools}
\begin{document}

\def\cl{\centerline}
\def\bd{\begin{description}}
\def\be{\begin{enumerate}}
\def\ben{\begin{equation}}
\def\benn{\begin{equation*}}
\def\een{\end{equation}}
\def\eenn{\end{equation*}}
\def\benr{\begin{eqnarray}}
\def\eenr{\end{eqnarray}}
\def\benrr{\begin{eqnarray*}}
\def\eenrr{\end{eqnarray*}}
\def\ed{\end{description}}
\def\ee{\end{enumerate}}
\def\al{\alpha}
\def\b{\beta}
\def\bR{\bar\R}
\def\bc{\begin{center}}
\def\ec{\end{center}}
\def\dg{\dagger}
\def\d{\dot}
\def\D{\Delta}
\def\del{\delta}
\def\ep{\epsilon}
\def\g{\gamma}
\def\G{\Gamma}
\def\h{\hat}
\def\iny{\infty}
\def\La{\Longrightarrow}
\def\la{\lambda}
\def\m{\mu}
\def\n{\nu}
\def\noi{\noindent}
\def\Om{\Omega}
\def\om{\omega}
\def\p{\psi}
\def\pr{\prime}
\def\r{\ref}
\def\R{{\bf R}}
\def\ra{\rightarrow}
\def\up{\uparrow}
\def\dn{\downarrow}
\def\lr{\leftrightarrow}
\def\s{\sum_{i=1}^n}
\def\si{\sigma}
\def\Si{\Sigma}
\def\t{\tau}
\def\th{\theta}
\def\Th{\Theta}

\def\vep{\varepsilon}
\def\vp{\varphi}
\def\pa{\partial}
\def\un{\underline}
\def\ov{\overline}
\def\fr{\frac}
\def\sq{\sqrt}
\def\ot{\otimes}
\def\tf{\textbf}
\def\WW{\begin{stack}{\circle \\ W}\end{stack}}
\def\ww{\begin{stack}{\circle \\ w}\end{stack}}
\def\st{\stackrel}
\def\Ra{\Rightarrow}
\def\R{{\mathbb R}}
\def\mf{\mathbf }
\def\bi{\begin{itemize}}
\def\ei{\end{itemize}}
\def\i{\item}
\def\bt{\begin{tabular}}
\def\et{\end{tabular}}
\def\lf{\leftarrow}
\def\nn{\nonumber}
\def\va{\vartheta}
\def\wh{\widehat}
\def\vs{\vspace}
\def\Lam{\Lambda}
\def\sm{\setminus}
\def\ba{\begin{array}}
\def\ea{\end{array}}
\def\bd{\begin{description}}
\def\ed{\end{description}}
\def\lan{\langle}
\def\ran{\rangle}
\def\l{\label}
\def\mb{\mathbb}
\def\ti{\times}
\def\v{\vec}
\large

\preprint{}
\title{ Tight lower bound to the geometric measure of quantum discord}

\date{\today}
\author{Ali Saif M. Hassan}
\email{alisaif73@gmail.com}
 \affiliation{Department of Physics, University of Amran, Amran, Yemen}
\author{Behzad Lari}
\email{behzadlari1979@yahoo.com}

\author{Pramod S. Joag}
\email{pramod@physics.unipune.ac.in}
\affiliation{Department of Physics, University of Pune, Pune, India-411007.}

\date{\today}
\begin{abstract}

Dakic, Vedral and Brukner [Physical Review Letters \tf{105},190502 (2010)] gave a geometric measure of quantum discord in a bipartite quantum state as the distance of the state from the closest classical quantum (or zero discord) state and derived an explicit formula for a two qubit state. Further, S.Luo and S.Fu [Physical Review A \tf{82}, 034302 (2010)] obtained a generic form of this geometric measure for a general bipartite state and established a lower bound. In this brief report we obtain a rigorous lower bound to the geometric measure of quantum discord in a general bipartite state which dominates that obtained by S.Luo and S.Fu.

\noi PACS numbers: 03.67.-a, 03.65.Ta
 \end{abstract}

\maketitle

Understanding quantum correlations in a multipartite quantum state is a fundamental open problem. Over the last two decades quantum correlations are studied in the entanglement-separability scenario \cite{horo} leading to important insights in quantum computing \cite{niel}, quantum communication protocols like teleportation \cite{benn,horo1}, superdense coding \cite{benn1}, cryptography \cite{gisi} etc. However, recently it was shown that  even some separable states contain nonclassical correlation and can be used to accomplish information processing tasks which cannot be achieved classically [7-13]. These nonclassical correlations of bipartite states are measured by quantum discord \cite{olli,hend}-the discrepancy between quantum versions of two classically equivalent expressions for mutual information. Quantum discord became a subject of intense research in different contexts [16-44]. As the evaluation of quantum discord involves optimization procedure, analytical results are known only in a few families of two-qubit states \cite{luo3,ali1,lang}. Recently, a necessary and sufficient condition for the existence of non-zero quantum discord was obtained and a geometrical way of quantifying quantum discord was proposed \cite{dakic}. Geometric measure of quantum discord, introduced by Dakic et al \cite{dakic}, is given by
\ben  \l{e1}
D(\rho)=\min_{\chi \in \Omega_0}||\rho-\chi||^2,
\een
where $\Omega_0$ denotes the set of zero-discord states and $||\rho-\chi||^2 := tr(\rho-\chi)^2$
is the square norm in the Hilbert-Schmidt space.  A state $\chi\in H^a\otimes H^b$ is of zero discord if and only if it is a classical-quantum state \cite{luo3,olli}, which can be represented as
\ben  \l{e2}
\chi = \sum_{k=1}^{m}p_k |k\ran\lan k|\otimes \rho_k ,
\een
where $\{p_k\}$ is a probability distribution, $\{|k\ran\}$ is an arbitrary orthonormal basis in $H^a$ and $\rho_k$ is a set of arbitrary states (density operators) acting on $H^b .$

 The quantum discord of a bipartite state $\rho$ on a system $H^a\otimes H^b$ with marginals $\rho^a$ and $\rho^b$ can be expressed as
\ben  \l{e4}
Q(\rho) = \min_{\Pi^a}
\{I(\rho)-I(\Pi^a(\rho))\}.
\een
Here the minimum is over von Neumann measurements (one dimensional
orthogonal projectors summing up to the identity) $\Pi^a$ $=\{\Pi_k^a\}$ on subsystem $a$, and
\ben
\Pi^a(\rho) = \sum_{k} (\Pi_k^a\otimes I^b)\rho (\Pi_k^a\otimes I^b)   \nn  \\
\een
is the resulting state after the measurement. $I (\rho) = S(\rho^a) + S(\rho^b)- S(\rho)$ is the quantum mutual information, $S(\rho) = -tr\rho\ln\rho$ is the von Neumann entropy, and $I^b$ is the identity operator on $H^b$.  Intuitively, quantum discord may thus be interpreted as the minimal loss of correlations (as
measured by the quantum mutual information) due to measurement. This formulation of quantum discord is equivalent to the original definition of quantum discord by Ollivier and Zurek \cite{olli}.

In order to obtain the desired lower bound for the quantum discord in an arbitrary bi-partite state, we set up the following scenario. Consider a bipartite system $H^a \otimes H^b$ with dim $H^a = m$ and dim $H^b = n$. Let $L(H^a)$ be the space consisting of all
linear operators on $H^a$. This is a Hilbert space with the Hilbert-Schmidt inner product
$$\langle X|Y \rangle = tr X^{\dagger} Y.$$
The Hilbert spaces $L(H^b)$ and $L(H^a\otimes H^b)$ are defined similarly.
Let $\{X_i : i = 1,2,\cdots,m^2\}$ and $\{Y_j : j = 1,2,\cdots,n^2\}$
be sets of Hermitian operators which constitute orthonormal bases for $L(H^a)$ and $L(H^b)$, respectively. Then
$$trX_i X_{i^{\prime}} = \delta_{ii^{\prime}} ,\;\; trY_jY_{j^{\prime}} = \delta_{jj^{\prime}}.$$
$\{X_i \otimes Y_j\}$ constitutes an orthonormal (product) basis for $L(H^a \otimes H^b)$ (linear operators on $H^a \otimes H^b$). In particular, any bipartite state $\rho$ on $H^a \otimes H^b$ can be expanded as
\ben   \l{e5}
\rho = \sum_{ij} c_{ij} X_i \otimes Y_j,
\een
with $c_{ij} = tr(\rho X_i \otimes Y_j).$

 S. Luo and S. Fu introduced the following form of geometric measure of quantum discord \cite{fu} (Here and throughout this article, the superscript $t$ denotes transpose of vectors or matrices).
\ben   \l{e6}
D(\rho) = tr(CC^t )- \max_{\h{A}} tr(\h{A}CC^t{\h{A}}^t ),
\een
where $C = [c_{ij} ]$ (Eq.(\r{e5})) is an $m^2\ti n^2$ matrix and the maximum is taken over all $m\times m^2$-dimensional isometric (see below) matrices $\h{A} = [a_{ki} ]$ such that
\begin{IEEEeqnarray}{rCl} \l{e7}
a_{ki} & = & tr(|k\rangle\langle k|X_i)=\lan k|X_i|k\ran,\;\; k = 1,2,\ldots,m  \nn  \\
   && i = 1,2,\ldots,m^2,
\end{IEEEeqnarray}
and $\{|k\rangle\}$ is any orthonormal basis in $H^a .$ We can expand the operator $|k\rangle\langle k|$ in this basis as
\begin{IEEEeqnarray}{rCl} \l{e8}
|k\rangle\langle k| & = & \sum_i a_{ki} X_i,\;\; k = 1,2,\ldots,m.
\end{IEEEeqnarray}
$\h{A}=[a_{ki}]$ is an isometry in the sense that $\h{A}{\h{A}}^t=I^a$ and the row vectors $\vec{a}_k$ of the matrix $\h{A}$ satisfy
\ben  \l{e9}
||\vec{a}_k||^2=\sum_{i=1}^{m^2} a_{ki}^2=1.
\een
Further, using their definition, it immediately follows that
\ben   \l{e10}
\sum_{k=1}^{m}a_{ki}=trX_i .
\een

We can represent the density operators acting on a bipartite system $H^a \otimes H^b ,$ with $dim H^a = m$
and $dim H^b = n$, as \cite{mahl,joag}
\begin{IEEEeqnarray}{rCl}  \l{e12}
\rho & = & \frac{1}{mn}(I_m\otimes I_n + \sum_i x_i \tilde{\lambda}_i\otimes I_n +\sum_j y_j I_m \otimes  \tilde{\lambda}_j \nn  \\
 && + \sum_{ij} t_{ij} \tilde{\lambda}_i \otimes \tilde{\lambda}_j),
\end{IEEEeqnarray}
where $\tilde{\lambda}_i , i=1,\ldots,m^2-1$ and $\tilde{\lambda}_j , j=1,\ldots,n^2-1$ are the generators of $SU(m)$ and $SU(n)$ respectively, satisfying $tr(\tilde{\lambda}_i \tilde{\lambda}_j)=2\delta_{ij}$  \cite{mahl}.
Notice that $\vec{x} \in \mathbb{R}^{m^2-1}$ and $\vec{y}\in \mathbb{R}^{n^2-1}$ are the coherence vectors of the subsystems $A$ and $B$, so that they can be determined locally. These are given by \cite{byrd,kim}
$$x_i=\frac{m}{2}tr(\rho \tilde{\lambda_i} \otimes I_n)=\frac{m}{2}tr(\rho_A \tilde{\lambda_i})$$
$$y_j=\frac{n}{2}tr(\rho I_m \otimes  \tilde{\lambda}_j )=\frac{n}{2}tr(\rho_B \tilde{\lambda_j}),$$
where $\rho_A=tr_B(\rho)$ and $\rho_B=tr_A(\rho)$ are the reduced density matrices. The correlation matrix $T=[t_{ij}]$ is given by
$$T=[t_{ij}]=\frac{m n}{4}[tr(\rho \tilde{\lambda_i} \otimes \tilde{\lambda_j})].$$

In this article, we find the lower bound of geometric measure of quantum discord which dominates the lower bound in ref. \cite{fu}.\\ 

\emph{Theorem 1}. Let $\rho$ be a bipartite state defined by Eq. (\r{e12}); then
\ben  \l{e13}
D(\rho) \geq \frac{2}{m^2n} (||\vec{x}||^2+\frac{2}{n} ||T||^2-\sum_{j=1}^{m-1} \eta_j),
\een

where $\eta_j,\;j=1,2,\cdots,m^2-1$ are the eigenvalues of the matrix $(\vec{x}\vec{x}^t+\frac{2TT^t}{n})$ arranged in non-increasing order (counting multiplicity).

We prove this theorem for arbitrary (finite) $m$ and $n$.

In Eq.(\r{e6}) giving the quantum discord $D(\rho),$ the maximum in the second term is taken over the $m\ti m^2$ isometric matrices $\h{A}$ which also satisfy Eq.s(\r{e7}) and Eq.(\r{e8}). In other words, the row vectors of $\h{A}$ are required to be the coherent vectors of states forming an orthonormal basis in $H^a .$ If we ignore this constraint while maximizing $tr(ACC^tA^t)$ and maximize over the isometric matrices $A$ defined below via Eq.(\r{eq2},\r{eq3}), the resulting maximum will be greater than or equal to the required maximum of $tr(\h{A}CC^t{\h{A}}^t )$ over the matrices $\h{A}$ satisfying Eq.s(\r{e7},\r{e8}). Since all the terms are positive, this leads to
\ben \l{eq1}
D(\rho) \geq  tr(CC^t )- \max_{A} tr(ACC^tA^t ).
\een

We proceed to obtain the maximum in the above inequality (Eq.(\r{eq1})). We choose the orthonormal bases $\{X_i\}$ and $\{Y_j\}$ in Eq.(\r{e5}) as the generators of $SU(m)$ and $SU(n)$ respectively \cite{mahl}.
\ben
X_1=\frac{1}{\sqrt{m}} I_m,  Y_1=\frac{1}{\sqrt{n}} I_n               \nn     \\
\een
 and
 \ben
 X_i=\frac{1}{\sqrt{2}} \tilde{\lambda}_{i-1},\; i=2,3,\ldots,m^2           \nn      \\
 \een
 \ben
 Y_j=\frac{1}{\sqrt{2}} \tilde{\lambda}_{j-1},\; j=2,3,\ldots,n^2.           \nn       \\
 \een

 Since $tr \tilde{\lambda}_i=0;\;i=1,2,\cdots,m^2-1$, we have, via Eq.(\r{e10}),
 \ben
 \sum_{k=1}^{m} a_{ki}=tr X_i = tr \tilde{\lambda}_i = 0 , i=2,\ldots,m^2.     \nn    \\
 \een
 Therefore,
 \ben  \l{e14}
 a_{mi}=-\sum_{k=1}^{m-1} a_{ki},\;i=2,3,\cdots,m^2.
\een
We now proceed to construct the $m\times m^2$ matrix $A$ defined via Eq.(\r{e7}-\r{e10}). We will use Eq.(\r{e14}). The row vectors of $A$ are
$$\vec{a}_k=(a_{k1},a_{k2},\cdots,a_{km^2}); k=1,2,\ldots,m.$$
Next we define
\ben   \l{e15}
\h{e}_k=\sqrt{\frac{m}{m-1}}(a_{k2},a_{k3},\ldots,a_{km^2}),\;k=1,2,\ldots,m-1
\een
and using Eq.(\r{e14}), we get
\ben   \l{e16}
\h{e}_m=-\sum_{k=1}^{m-1} \h{e}_k.
\een
We can prove
\ben  \l{e17}
||\h{e}_k||^2=1\;\;k=1,2,\ldots,m-1
\een
 using the condition $||\vec{a}_k||^2=\sum_{i=1}^{m^2} a_{ki}^2=1$ (Eq.(\r{e9})) and using Eq.(\r{e10}) with $i=1,$ namely, $a_{k1}= tr(|k\rangle \langle k| X_1)=\frac{1}{\sqrt{m}}.$ Further, isometry of the $A$ matrix ($AA^{t}=I$) implies
\ben  \l{e18}
\h{e}_i{\h{e}_j}^{t}=\frac{-1}{m-1}, j\ne i=1,2,\ldots,m-1.
\een
We can now construct the row vectors of $m\times m^2$ matrix $A$, using Eq.(\r{e15}) and Eq.(\r{e16}),
\ben \l{eq2}
\vec{a}_k=\frac{1}{\sqrt{m}} (1, \sqrt{m-1}\h{e}_k),\; k=1,2,\cdots,m-1        \\
\een
\ben  \l{eq3}
\vec{a}_m=\frac{1}{\sqrt{m}} (1, - \sqrt{m-1} \sum_{k=1}^{m-1} \h{e}_k)             \\
\een
defining matrix $A.$ 

We get the elements of $C=[c_{ij}]=[tr(\rho X_i\otimes Y_j)]$ using the definitions of the bases $\{X_i\}$ and $\{Y_j\}$ given above, in terms of the generators of $SU(m)$ and $SU(n)$. This gives
\begin{displaymath}
C =
\left(\begin{array}{cc}
\frac{1}{\sqrt{mn}} & \frac{\sqrt{2}}{n\sqrt{m}}\vec{y}^t \\
\frac{\sqrt{2}}{m\sqrt{n}}\vec{x} & \frac{2}{mn} T\\
\end{array}\right),
\end{displaymath}
 and
 \begin{IEEEeqnarray}{rCl} \l{e19}
 tr(CC^t) & = & (\frac{1}{mn}+\frac{2}{n^2 m} ||\vec{y}||^2+\frac{2}{m^2n} ||\vec{x}||^2  \nn  \\
 &&+\frac{4}{n^2m^2} ||T||^2).
\end{IEEEeqnarray}
where $\v{x},$ $\v{y}$ and $T$ are the coherent vectors and the correlation matrix respectively, defined in Eq.(\r{e12}). 

Having constructed the matrices $A$ and $C,$ we get, for $tr(ACC^tA^t),$
\begin{widetext}
\ben  \l{e20}
tr(ACC^tA^t) = \frac{1}{m}\left\{\frac{1}{n}+\frac{2}{n^2}||\vec{y}||^2 +\frac{2(m-1)}{m^2n}\left[\sum_{j=1}^{m-1}\h{e}_jG\h{e}_j^t +
\sum_{i=1}^{m-1}\sum_{j=1}^{m-1}\h{e}_i G\h{e}_j^t \right]\right\},
\een
\end{widetext}
where
\ben  \l{e21}
G=\vec{x}\vec{x}^t + \frac{2TT^t}{n}
\een
is the $(m^2-1)\ti (m^2-1)$ real symmetric matrix. The eigenvectors of $G$ span $\mb{R}^{m^2-1}$ and form a orthonormal basis of $\mb{R}^{m^2-1}.$ Let $\eta_1 ,\eta_2 ,\ldots,\eta_{m^2-1}$ be the eigenvalues  of $G$ arranged in non-increasing order (counting multiplicity). Let $(|\h{f}_1\rangle ,|\h{f}_2\rangle ,\ldots,|\h{f}_{m^2-1}\rangle)$ be the corresponding orthonormal eigenvectors of $G.$  To maximize the right hand side of Eq.(\r{e20}), we choose
\ben  \l{e22}
\h{e}_1= \h{f}_1
\een
and expand $\{\h{e}\}_{j=2}^{m-1}$ in the eigenbasis of $G.$ Thus,
\ben  \l{e23}
\h{e}_j = \sum_{k=1}^{m^2-1}\ep^{(j)}_k \h{f}_k\;\; j=2,3,\ldots,m-1.
\een
Eq.(\r{e17}) with $\h{e}_i$ replaced by $\h{e}_1$ gives us
\benr  \l{e24}
\ep^{(1)}_1 & = & 1   \nn   \\
\ep^{(j)}_1 & = & \frac{-1}{m-1}, j=2,3,\ldots,m-1.
\eenr
Since $\h{e}_1= \h{f}_1$ (Eq.(\r{e22})) and $||\h{e}_j||^2=1$ for all $j$ (Eq.(\r{e17})) we get
\ben  \l{e25}
 \sum_{k=1}^{m^2-1}{\ep^{(j)}_k}^2=1.
\een
We substitute Eq.(\r{e23}) for $\h{e}_j$ in the expression for $tr(ACC^tA^t)$ (Eq.(\r{e20})) and use Eq.(\r{e24}) to get
\begin{widetext}
\begin{IEEEeqnarray}{rCl} \l{e26}
tr(ACC^tA^t) & = &  \frac{1}{m}\left\{\frac{1}{n}+\frac{2}{n^2}||\vec{y}||^2+
 \frac{2(m-1)}{m^2n}\left[\frac{m}{m-1}\eta_1
   + \: 2\sum_{j=2}^{m-1}\sum_{k=2}^{m^2-1}{\ep^{(j)}_k}^2\eta_k \right.\right.   \nn  \\
  && + \: \left.\left. 2\sum_{i=2}^{m-2}\sum_{j>i}^{m-1}\sum_{k=2}^{m^2-1}\ep^{(i)}_k\ep^{(j)}_k \eta_k\right]\right\}.
\end{IEEEeqnarray}
\end{widetext}

We have to choose vectors $\{\h{e}_j \}\;j=1,\ldots,m-1 ,$ that is, the expansion coefficients $\{\ep^{(j)}_k\} ;\;j=1,\ldots,m-1:\;k=1,\ldots,m^2-1$ in Eq.(\r{e23}), consistent with Eq.s(\r{e24},\r{e25}), so as to maximize $tr(ACC^tA^t).$ First we note that we must use at least $(m-1)$ eigenvectors $\h{f}_k$ in Eq.(\r{e23}) to expand all of $\{\h{e}_j \}\;j=1,\ldots,m-1 ,$ because otherwise $\{\h{e}_j \}$ becomes a set of $m-1$ linearly dependent vectors in a subspace of dimension less than $m-1 ,$ in which case the row vectors of matrix $A$ cease to be mutually orthogonal. Thus, for $m=3$ a choice like $\h{e}_2=\ep^{(2)}_1 \h{f}_1$ yields $\vec{a}_1\cdot\vec{a}_2 \neq 0 .$ With eigenvalues $\{\eta_k\}\;k=1,\ldots,m^2-1$ arranged in non-increasing order, we now see that, for every choice of $\{\ep^{(j)}_k\} ;\;j=1,\ldots,m-1;\;k=1,\ldots,m^2-1$ we get the following upper bound on $tr(ACC^tA^t)$
\begin{widetext}
\begin{IEEEeqnarray}{rCl} \l{e27}
tr(ACC^tA^t) & \le & \frac{1}{m}\left\{\frac{1}{n}+\frac{2}{n^2}||\vec{y}||^2+
 \frac{2(m-1)}{m^2n}\left[\frac{m}{m-1}\eta_1 +  2\sum_{j=2}^{m-1}\left(\sum_{k=2}^{j-1}{\ep^{(j)}_k}^2\eta_k + \eta_{j}\sum_{k=j}^{m^2-1}{\ep^{(j)}_k}^2\right)\right.\right. \nn  \\
 && \left.\left. +2\sum_{i=2}^{m-2}\sum_{j>i}^{m-1}\left(\sum_{k=2}^{i-1}\ep^{(i)}_k\ep^{(j)}_k \eta_k + \eta_{i}\sum_{k=i}^{m^2-1}\ep^{(i)}_k\ep^{(j)}_k \right)\right]\right\}.
\end{IEEEeqnarray}
\end{widetext}
The maximum value of RHS is then obtained by choosing $\ep^{(j)}_k=0$ for all $k>j ,$ that is,
\ben \l{e28}
\h{e}_j = \sum_{k=1}^{j}\ep^{(j)}_k\h{f}_k\;\; j=2,3,\ldots,m-1,
\een
 which leads, using the fact that $\ep^{(j)}_k = 0$ if $k>j ,$ to
\begin{widetext}
 \begin{IEEEeqnarray}{rCl} \l{e30}
\max_{A}[tr(ACC^tA^t)] & = & \frac{1}{m}\left\{\frac{1}{n}+\frac{2}{n^2}||\vec{y}||^2+
 \frac{2(m-1)}{m^2n}\left[\frac{m}{m-1}\eta_1
 + 2\sum_{k=2}^{m-1}\eta_k\left[\sum_{j=k}^{m-1}{\ep^{(j)}_k}^2 \right.\right.\right. \nn  \\
&&\left.\left.\left. + \sum_{i=k}^{m-2}\sum_{j>i}^{m-1}\ep^{(i)}_k\ep^{(j)}_k \right]\right]\right\}.
\end{IEEEeqnarray}
\end{widetext}

The remaining task is to find the coefficients $\ep^{(j)}_k ,\;k=1,2,\ldots,j$ in the expansion of $\h{e}_j$ in the eigenbasis $\{\h{f}_k\},\;k=1,2,\ldots,j$ which satisfy Eq.s(\r{e16},\r{e17},\r{e28}) and the consequential equations (\r{e24},\r{e25}). Using Eq.s(\r{e17},\r{e18},\r{e24},\r{e25},\r{e28}) the problem can be reduced to the coupled pair of equations 
\ben  \l{e34}
{\ep^{(j)}_j}^2=\frac{m(m-2)}{(m-1)^2}-\sum_{k=2}^{j-1}{\ep^{(j)}_k}^2\; ; j>1
\een
\ben \l{e35}
\ep^{(j)}_i = \frac{1}{\ep^{(i)}_i}\left[\frac{-m}{(m-1)^2}-\sum_{k=2}^{i-1}\ep^{(i)}_k\ep^{(j)}_k\right]\; ; j>i>1.
\een
which can be solved iteratively, starting from $j=2 .$ The result is
\ben \l{e42}
{\ep^{(j)}_{j}}^2=\frac{m}{(m-1)}\left[\frac{m-j}{m-j+1}\right],
\een
and
\ben \l{e43}
\ep^{(j)}_i = \frac{-\sq{m}}{\sq{(m-1)(m-i+1)(m-i)}}  \; ; j>i
\een
Using Eq.s(\r{e42},\r{e43}) the last term in Eq.(\r{e30}) can be evaluated. We have,
\ben \l{e44}
\sum_{j=k}^{m-1}{\ep^{(j)}_k}^2 + \sum_{i=k}^{m-2}\sum_{j>i}^{m-1}\ep^{(i)}_k\ep^{(j)}_k = \frac{m}{2(m-1)}.
\een
 Eq.(\r{e44}) simplifies Eq.(\r{e30}) to
 \ben \l{e45}
 \max_A [tr(ACC^tA^t)] = \frac{1}{mn}+\frac{2}{n^2m}||\vec{y}||^2 +\frac{2}{m^2 n}\sum_{j=1}^{m-1}\eta_j
 \een
 Finally, Eq.(\r{e19}), Eq.(\r{e45}) and Eq.(\r{eq1}) together imply
 \ben
 D(\rho) \geq \frac{2}{m^2 n} (||\vec{x}||^2+\frac{2}{n} ||T||^2-\sum_{j=1}^{m-1} \eta_j),   \nn  \\
\een
which completes the proof of the theorem.

Now, we prove that 
\ben \l{e46}
D(\rho) \geq \frac{2}{m^2 n} (||\vec{x}||^2+\frac{2}{n} ||T||^2-\sum_{j=1}^{m-1} \eta_j^{\downarrow}) \geq tr(CC^t)-\sum_{i=1}^m \lambda_i^{\downarrow}
\een
where $\lambda_i^{\downarrow}$ are the eigenvalues of $CC^t$ listed in decreasing order (counting multiplicity) and $\eta_j^{\downarrow}$ are the eigenvalues of $G=\vec{x}\vec{x}^t + \frac{2TT^t}{n}$ listed in decreasing order (counting multiplicity). The last lower bound in Eq.(\r{e46}) is proved in ref. \cite{fu} where the inequality $\max_{\h{A}} tr(\h{A}CC^t{\h{A}}^t) \le \sum_{i=1}^m \lambda_i^{\downarrow}$ is derived. Since we have proved $\max_{\h{A}} tr(\h{A}CC^t{\h{A}}^t) \le \frac{1}{m n}+\frac{2}{n^2 m}||\v{y}||^2+\frac{2}{m^2 n}\sum_{i=1}^{m-1} \eta_i^{\downarrow}= a+\frac{2}{m^2 n}\sum_{i=1}^{m-1} \eta_i^{\downarrow},$ to prove Eq.(\r{e46}), it is enough to prove 
\ben \l{e47}
\sum_{i=1}^m \lambda_i^{\downarrow} \ge a+\frac{2}{m^2 n}\sum_{i=1}^{m-1} \eta_i^{\downarrow}.
\een
Let us rewrite $CC^t$ as
\begin{displaymath}
CC^t =
\left(\begin{array}{cc}
a & \v{u}^t \\
\v{u} & \frac{2}{m^2n} G\\
\end{array}\right),
\end{displaymath}
where $a=\frac{1}{m n}+\frac{2}{n^2 m}||\v{y}||^2$ , $G=\vec{x}\vec{x}^t + \frac{2TT^t}{n}$ and $\v{u}=\frac{\sqrt{2}}{m n \sqrt{m}}\v{x}+\frac{2\sqrt{2}}{m n^2 \sqrt{m}} T \v{y}.$
Let eigenvalues of $\frac{2}{m^2n} G$ be $\{\eta^{\prime}_j\}=\frac{2}{m^2 n}\{\eta_j\}$, where $ \{\eta_j\}$ are the eigenvalues of $G$. Then
\ben \l{e48}
tr(CC^t)= a+\frac{2}{m^2 n} \sum_{j=1}^{m^2-1}\eta_j=\sum_{i=1}^{m^2} \lambda_i.
\een
Let $\lambda^{\uparrow}_1 \le \lambda^{\uparrow}_2 \le \ldots \le \lambda^{\uparrow}_{m^2}$ and $\eta^{\prime \uparrow}_1 \le \eta^{\prime \uparrow}_2 \le \ldots \le \eta^{\prime \uparrow}_{m^2-1}.$ From theorem (4.3.8) in \cite{horn}, we have
 $$ \lambda^{\uparrow}_1 \le \eta^{\prime \uparrow}_1 \le \lambda^{\uparrow}_2 \le \eta^{\prime \uparrow}_2 \le \ldots \le \eta^{\prime \uparrow}_{m^2-1} \le \lambda^{\uparrow}_{m^2},$$ so that
 \ben \l{e49}
 \sum_{i=1}^{m^2-m}\lambda^{\uparrow}_i \le \sum_{j=1}^{m^2-m} \eta^{\prime \uparrow}_j .
 \een
Now, we use Eq.(\r{e48}) to get
\ben \l{e50}
  \sum_{i=1}^{m^2-m} \lambda^{\uparrow}_i+\sum_{i=m^2-m+1}^{m^2} \lambda^{\uparrow}_i= a+ \sum_{j=1}^{m^2-m}\eta^{\prime \uparrow}_j+ \sum_{j=m^2-m+1}^{m^2-1}\eta^{\prime \uparrow}_j.
\een
Replacing the first term on the left hand side of Eq.(\r{e50}) by $\sum_{j=1}^{m^2-m} \eta^{\prime \uparrow}_j$ we get, using Eq.(\r{e49}),
\ben \l{e51}
 \sum_{i=m^2-m+1}^{m^2}\lambda^{\uparrow}_i \ge \sum_{j=m^2-m+1}^{m^2-1} \eta^{\prime \uparrow}_j+a,
 \een
 Finally, we use $ \lambda^{\uparrow}_j=\lambda^{\downarrow}_{m^2-j+1}$ and $ \eta^{\prime \uparrow}_j =\eta^{\prime \downarrow}_{m^2-j}$ \cite{bhat}, to get
 \ben
 \sum_{i=1}^{m}\lambda^{\downarrow}_i \ge \sum_{j=1}^{m-1} \eta^{\prime \downarrow}_j+a . \nn
 \een

\emph{Examples}

(1) We consider the two qutrit state
\ben \l{e52}
\rho=p |e\rangle\langle e|+ (1-p)\frac{I}{9} \\
\een
where $|e\rangle =\frac{1}{\sqrt{6}}(|2\rangle\otimes |2\rangle +|3\rangle\otimes |3\rangle +|2\rangle\otimes |1\rangle +|1\rangle\otimes |2\rangle +|1\rangle\otimes |3\rangle +|3\rangle\otimes |1\rangle) $, $I$ is the identity operator and $\{|i\rangle; i=1,2,3\}$ is the standard basis in $C^3$.
Fig. 1 shows the variation of lower bound (given in Eq.(\r{e13})) and the lower bound on $D(\rho),$ as given in \cite{fu}, namely, $tr(CC^{t})-\sum_{i=1}^{m}\lambda_{i}$ (where $\{\lambda_{i}\}$ are the eigenvalues of $CC^{t}$ listed in the decreasing order, counting multiplicity), with $p.$ 
 \begin{figure}[!ht]
\begin{center}
\includegraphics[width=8cm,height=5cm]{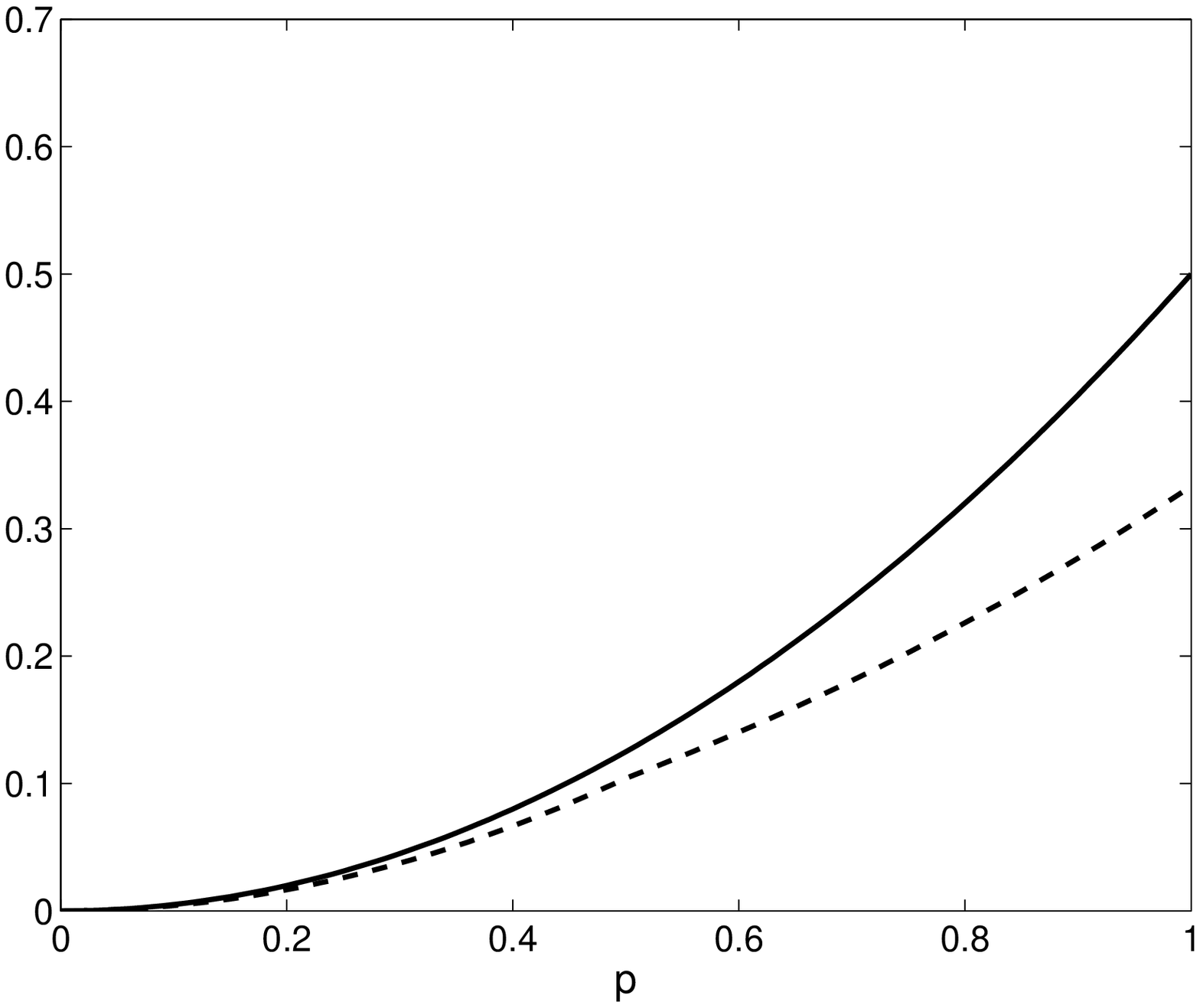}\\
Fig. 1 Lower bound of quantum discord (Eq.(\r{e13}), solid line) and lower bound (Eq.(6) in \cite{fu}, dashed line) as a function of $p$ (Eq.(\r{e52})).
\end{center}
\end{figure}\\
We see that the lower bound in Eq.(\r{e13}) dominates this lower bound for $p > 0.2.$

(2) We consider the two qutrit state
\ben \l{e53}
\rho=p |e_1\rangle\langle e_1|+ (1-p)|e_2\rangle\langle e_2| \\
\een
where  $|e_1\rangle =\frac{1}{2} |11\rangle+\frac{1}{2} |22\rangle+\frac{1}{\sqrt{2}} |33\rangle$ and $|e_2\rangle =\frac{1}{\sqrt{6}}(|2\rangle\otimes |2\rangle +|3\rangle\otimes |3\rangle +|2\rangle\otimes |1\rangle +|1\rangle\otimes |2\rangle +|1\rangle\otimes |3\rangle +|3\rangle\otimes |1\rangle).$ Fig. 2 shows the variation of $D(\rho)$ (given in Eq.(\r{e13}) and the lower bound on $D(\rho),$ as given in \cite{fu}, namely, $tr(CC^{t})-\sum_{i=1}^{m}\lambda_{i}$ (where $\{\lambda_{i}\}$ are the eigenvalues of $CC^{t}$ listed in the decreasing order, counting multiplicity), with $p.$
\begin{figure}[!ht]
 \begin{center}
\includegraphics[width=8cm,height=5cm]{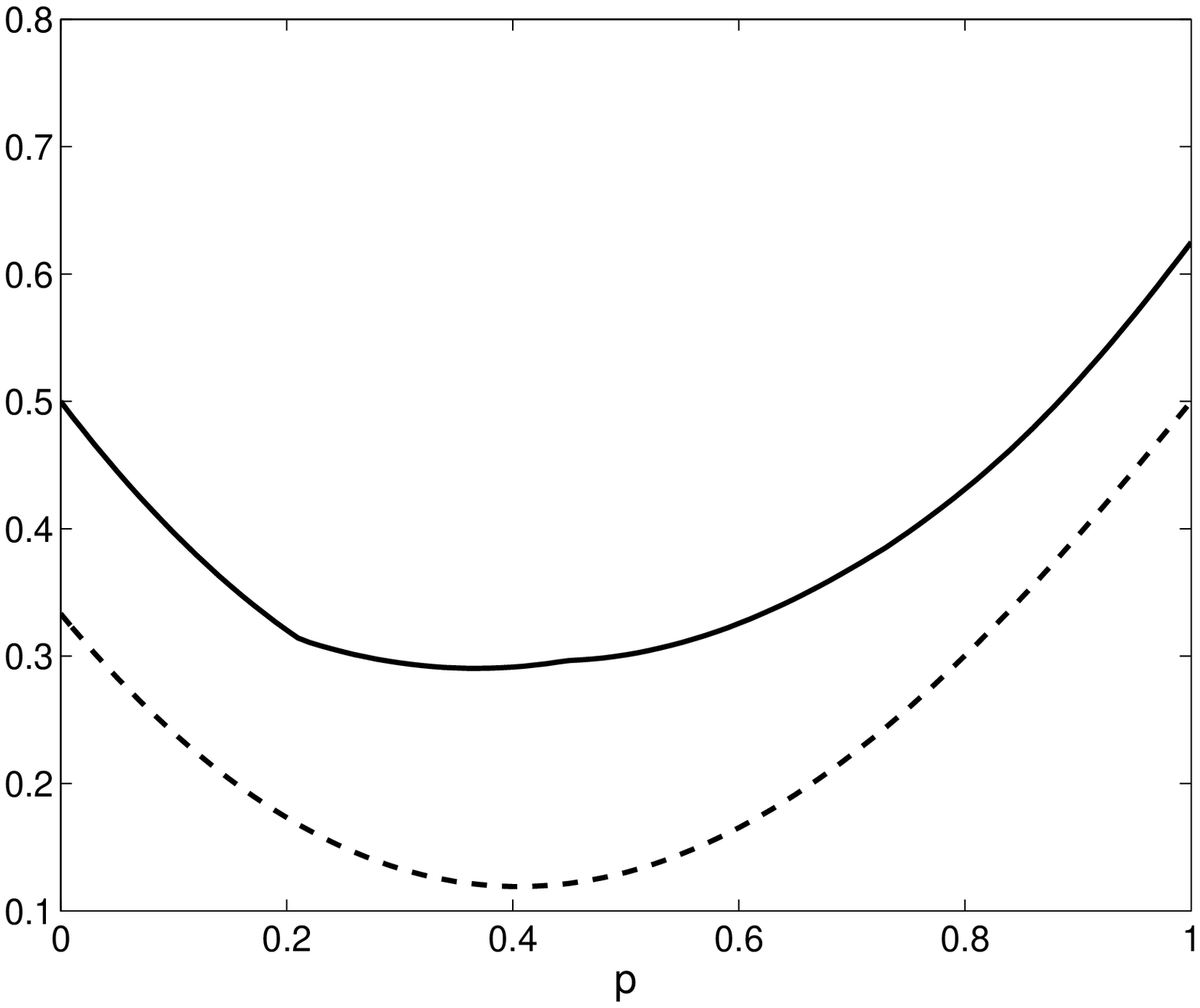}\\
Fig. 2 Quantum discord (Eq.(\r{e13}), solid line) and lower bound (Eq.(6) in \cite{fu}, dashed line) as a function of $p$ (Eq.(\r{e53})).
\end{center}
\end{figure}\\
We see that the lower bound in Eq.(\r{e13}) dominates this lower bound.

To the best of our knowledge, only the lower bounds on the quantum discord in an arbitrary bipartite state are presently available. This seems quite surprising as detecting quantum discord is not a NP-hard problem such as the separability problem and one expects an exact computable expression for quantum discord in all dimensions. This expectation is further augmented by the fact that the set of zero discord states is of measure zero. In order to get such an exact computable expression for quantum discord in all dimensions, we have to maximize the second term in Eq.(\r{e6}) over the set of $m\ti m^2$ matrices $A$ which satisfy (in addition to other conditions) the requirement that each row vector of $A$ must be a coherent vector of a orthonormal basis state in $H^a .$ For the two qubit case, this requirement becomes redundant as every unit vector in ${\mb{R}}^3$ is a coherent vector of some single qubit pure state. If $dim(H^a) > 2,$ not every unit vector in ${\mb{R}}^{d^2 -1}$ is a coherent vector of some pure state in $H^a .$ Hence whenever $dim(H^a) > 2,$ the above requirement is to be included as an independent constraint in the constrained optimization of the second term in Eq.(\r{e6}). The resulting constrined optimization problem is very difficult because the set of coherent vectors (for $dim(H^a) > 2$) do not have some simple geometric structure like Bloch sphere. This is the reason why an exact computable expression for quantum discord for all dimensions still eludes us. We note that the lower bound in Eq.(\r{e13}) becomes exact for a $2\ti d$ system (with measurement on the qubit) since the above constraint is relaxed in this case. Finally, our lower bound on discord can be meaningfully used to compare discordant states and check on the possible monogamy property of dicord \cite{gio}.        

\emph{Acknowledgments }:\\
This work was supported by the BCUD research grant RG-13. ASMH  thanks Pune University for the hospitality during his visit when this work was initiated. We thank the anonymous referee whose suggestions have contributed towards the improvement of this paper.


\begin{thebibliography}{99}

\bibitem{horo}
  R. Horodecki, P. Horodecki, M. Horodecki, and K. Horodecki, Rev. Mod. Phys. \textbf{81}, 865 (2009).
 \bibitem{niel}
 M. A. Nielsen and I. L. Chuang, Quantum Computation and Quantum Information, (Cambridge University Press, Cambridge, England, 2000).
\bibitem{benn}
C. H. Bennett et. al., Phys. Rev. Lett. \textbf{70}, 1895 (1993).
\bibitem{horo1}
 R. Horodecki, P. Horodecki and M. Horodecki, Phys. Lett. A \textbf{200}, 340 (1995).
\bibitem{benn1}
 C. H. Bennett and S. J. Wiesner, Phys. Rev. Lett. \textbf{69}, 2881 (1992).
\bibitem{gisi}
 N. Gisin, et. al. , Rev. Mod. Phys. \textbf{74}, 145 (2002).
\bibitem{knil}
 E. Knill and R. Laflamme, Phys. Rev. Lett. \textbf{81}, 5672 (1998).
 \bibitem{brau}
 S. L. Braunstein, C. M. Caves, R. Jozsa, N. Linden,  S. Popescu, and R. Schack, Phys. Rev. Lett. \textbf{83}, 1054 (1999).
\bibitem{meye}
 D. A. Meyer, Phys. Rev. Lett. \textbf{85}, 2014 (2000).
\bibitem{datt}
A. Datta, S. T. Flammia, and C. M. Caves, Phys. Rev. A \textbf{72}, 042316 (2005).
\bibitem{datt1}
 A. Datta and G. Vidal, Phys. Rev. A \textbf{75}, 042310 (2007).
\bibitem{datt2}
 A. Datta, A. Shaji, and C. M. Caves, Phys. Rev. Lett. \textbf{100}, 050502 (2008).
 \bibitem{lany}
 B. P. Lanyon, M. Barbieri, M. P. Almeida, and A. G. White, Phys. Rev. Lett. \textbf{101}, 200501 (2008).
\bibitem{olli}
H. Ollivier and W. H. Zurek, Phys. Rev. Lett. \textbf{88}, 017901 (2001).
\bibitem{hend}
 L. Henderson and V. Vedral, J. Phys. A \textbf{34}, 6899 (2001).
\bibitem{zure}
 W. H. Zurek, Phys. Rev. A \textbf{67}, 012320 (2003).
\bibitem{zhan}
 Z. Zhang and S. Luo, Phys. Rev. A \textbf{75}, 032312 (2007).
\bibitem{luo3}
 S. Luo, Phys. Rev. A \textbf{77}, 042303 (2008).
\bibitem{dill}
 R. Dillenschneider, Phys. Rev. B \textbf{78}, 224413 (2008).
\bibitem{shab}
 A. Shabani and D. A. Lidar, Phys. Rev. Lett. \textbf{102}, 100402 (2009).
\bibitem{datt3}
 A. Datta and S. Gharibian, Phys. Rev. A 79, 042325 (2009).
\bibitem{sara}
 M. S. Sarandy, Phys. Rev. A 80, 022108 (2009).
\bibitem{werl1}
 T.Werlang, S. Souza, F. F. Fanchini, and C. J. Villas Boas, Phys. Rev. A 80, 024103 (2009).
\bibitem{datt4}
 A. Datta, Phys. Rev. A 80, 052304 (2009).
\bibitem{wang}
 B. Wang, Z. Y. Xu, Z. Q. Chen, and M. Feng, Phys. Rev. A 81, 014101 (2010).
\bibitem{modi}
 K. Modi, T. Paterek, W. Son, V. Vedral, and M. Williamson, Phys. Rev. Lett. 104, 080501 (2010).
\bibitem{chen}
 Y. X. Chen and S. W. Li, Phys. Rev. A 81, 032120 (2010).
\bibitem{werl2}
 T. Werlang and G. Rigolin, Phys. Rev. A 81, 044101 (2010).
\bibitem{ali1}
 M. Ali, A. R. P. Rau, and G. Alber, Phys. Rev. A 81, 042105 (2010).
\bibitem{fanc}
 F. F. Fanchini, T. Werlang, C. A. Brasil, L. G. E. Arruda, and A. O. Caldeira, Phys. Rev. A 81, 052107 (2010).
\bibitem{ferr}
 A. Ferraro, L. Aolita, D. Cavalcanti, F. M. Cucchietti, and A. Ac´ýn, Phys. Rev. A 81, 052318 (2010).
\bibitem{mazz}
 L. Mazzola, J. Piilo, and S. Maniscalco, Phys. Rev. Lett. 104, 200401 (2010).
\bibitem{wang2}
 J. Wang, J. Deng, and J. Jing, Phys. Rev. A 81, 052120 (2010).
\bibitem{byli}
 B. Bylicka and D. Chruscinski, Phys. Rev. A 81, 062102 (2010).
\bibitem{brod}
 A. Brodutch and D. R. Terno, Phys. Rev. A 81, 062103 (2010).
\bibitem{zwol}
 M. Zwolak, H. T. Quan, and W. H. Zurek, Phys. Rev. A 81,
062110 (2010).
\bibitem{soar}
 D. O. Soares-Pinto, L. C. C´eleri, R. Auccaise, F. F. Fanchini, E. R. deAzevedo, J. Maziero, T. J. Bonagamba, and R. M. Serra, Phys. Rev. A 81, 062118 (2010).
\bibitem{gege}
 R. C. Ge, M. Gong, C. F. Li, J. S. Xu, and G. C. Guo, Phys. Rev. A 81, 064103 (2010).
\bibitem{giod}
 P. Giorda and M. G. A. Paris, Phys. Rev. Lett. 105, 020503 (2010).
\bibitem{vasi}
 R. Vasile, P. Giorda, S. Olivares, M. G. A. Paris, and S. Maniscalco, Phys. Rev. A 82, 012313 (2010).
\bibitem{ades}
 G. Adesso and A. Datta, Phys. Rev. Lett. 105, 030501 (2010).
\bibitem{xuxu}
 J.-S. Xu, X.-Y. Xu, C.-F. Li, C.-J. Zhang, X.-B. Zou, and G.-C. Guo, Nature Commun. 1, 7 (2010).
\bibitem{cuii}
 J. Cui and H. Fan, J. Phys. A 43, 045305 (2010).
\bibitem{lang}
 M. D. Lang and C. M. Caves, Phys. Rev. Lett. 105, 150501 (2010).
 \bibitem{dakic}
 B. Dakic, V. Vedral, and C. Brukner, Phys. Rev. Lett. 105,190502 (2010).
\bibitem{fu}
 Shunlong Luo and Shuangshuang Fu, Phys. Rev. A \textbf{82}, 034302 (2010).
\bibitem{mahl}
G. Mahler, Volker A. Weberruss, \emph{Quantum Networks} (Springer-Verlag Berlin Heidelberg 1995).
\bibitem{joag}
Ali Saif M. Hassan and Pramod S. Joag, Physical Review A \textbf{77}, 062334 (2008).
\bibitem{byrd}
M. S. Byrd and N. Khaneja, Phys. Rev. A \textbf{68}, 062322 (2003); Ali Saif M. Hassan and Pramod S. Joag, Quantum Inf. Comput.
\textbf{8}, 773 (2008).
\bibitem{kim}
 G. Kimura, Phys. Lett. A \textbf{314}, 339 (2003).
\bibitem{horn}
 R. Horn  and C. Johnson,  \emph{Matrix Analysis} (Cambridge University Press 1990 ).
\bibitem{bhat}
 R. Bhatia,  \emph{Matrix Analysis} (Springer-Verlag New York, Inc. 1997), chapter II, page 28 Eq.(II.1).
\bibitem{gio}
G. L. Giorgi, Phys. Rev. A 84, 054301 (2011).
\end{thebibliography}
\end{document}